\documentclass[twocolumn,showpacs,pre]{revtex4}
\usepackage{dcolumn}
\usepackage{graphicx}
\usepackage{graphicx,amssymb,amsmath,revsymb}
\usepackage{natbib}

\usepackage{tikz}
\usetikzlibrary{arrows}

\begin{document}

\title{Local mechanical response in semiflexible polymer networks subjected to an axisymmetric prestress}

\author{David A. Head$^{1}$}
\email{d.head@leeds.ac.uk}

\author{Daisuke Mizuno$^{2}$}

\affiliation{$^{1}$School of Computing, Leeds University, Leeds LS2 9JT, United Kingdom}

\affiliation{$^{2}$Department of Physics, Faculty of Exact Sciences, Kyushu University, Fukuoka 812-8581, Japan}

\date{\today}

\begin{abstract}
Analytical and numerical calculations are presented for the mechanical response of fiber networks in a state of axisymmetric prestress, in the limit where geometric non-linearities such as fiber rotation are negligible. This allows us to focus on the anisotropy deriving purely from the non-linear force-extension curves of individual fibers. The number of independent elastic coefficients for isotropic, axisymmetric and fully anisotropic networks are enumerated, before deriving expressions for the response to a locally applied force that can be tested against {\em e.g.} microrheology experiments. Localised forces can generate anisotropy away from the point of application, so numerical integration of non-linear continuum equations is employed to determine the stress field, and induced mechanical anisotropy, at points located directly behind and in front of a force monopole. Results are presented for the wormlike chain model in normalised forms, allowing them to be easily mapped to a range of systems. Finally, the relevance of these findings to naturally occurring systems and directions for future investigation are discussed.
\end{abstract}

\pacs{87.16.Ka, 46.15.Cc, 46.25.Cc}

\maketitle

\section{Introduction}

Many materials of industrial and biological importance are fibrous in nature, including paper~\cite{Alava2006}, carbon nanotube assemblies~\cite{Hall2008}, and a range of protein fiber networks such as the eukaryotic cellular cytoskeleton~\cite{HowardBook,BrayBook,AlbertsBook}, some pathological and functional amyloids~\cite{Xu2010,Knowles2011,Ndlovu2012,Cabriolu2012}, self-assembled peptide networks~\cite{Helen2011,Raeburn2012,Roberts2012,Tang2011} and some scaffolds used in tissue engineering~\cite{Raif2008,Mujeeb2012,HainesButterick2007,BurdickMauck}. There now exists a substantial literature relating the macroscopic viscoelastic properties of such networks to their underlying microscopic architecture~\cite{Gittes1998,Morse1998,Head2003a,Head2003b} that has been verified for actin networks in particular~\cite{Gardel2004,Storm2005}, but also other protein fibers such as vimentin~\cite{Lin2010}. This theoretical framework is however not exhaustive, and noticeably the issue of anisotropy has received little attention to date, despite visualization of the actin cortex frequently demonstrating a preferred orientation~\cite{BrayBook,AlbertsBook,Shutova2012}. This {\em morphological} anisotropy has been investigated in the context of coarse graining, highlighting relevant perturbations to the isotropic case~\cite{Missel2010}, and coupling to the environment leading to the alignment of stress fibers~\cite{Zemel2010}.

\begin{figure}[htbp]
\centerline{\includegraphics[width=8cm]{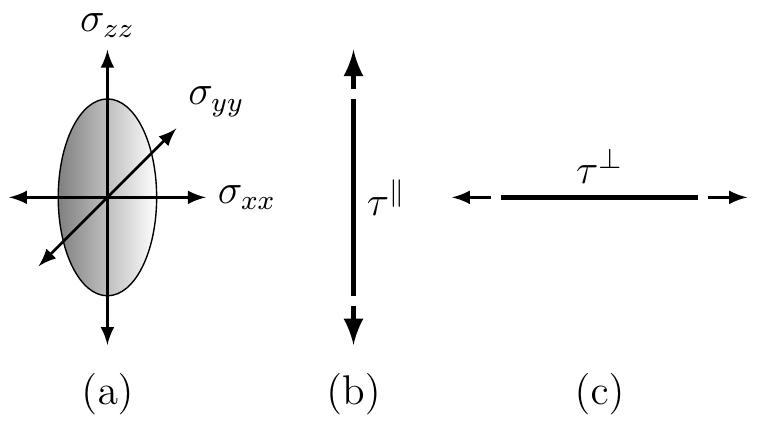}} 
\caption{Schematic of prestress-induced anisotropy. (a) The axes of principal stresses define Cartesian coordinates with the $z$-axis aligned with the maximum stress. (b)~Fibers aligned parallel to this axis are placed in a tension~$\tau^{\parallel}$. (c)~Those in the transverse direction have a tension~$\tau^{\perp}$, where $\tau^{\perp}<\tau^{\parallel}$. If the force-extension curve is non-linear, the network will thus become mechanically anisotropic, while potentially remaining (to good approximation) geometrically isotropic.}
\label{f:schematic}
\end{figure}

However, there is another form of mechanical anisotropy that arises, not from the geometric microstructure of the network, but rather from the non-linear response of individual fibers. Consider applying an anisotropic prestress to an isotropic fiber network. This prestress can emerge spontaneously due to intracellular mechanisms~\cite{Park2010}, or from the sustained uniaxial strains employed to check the non-linear properties of nanofiber scaffolds in tissue engineering~\cite{Meng2012,Tronci2013}, for example. If the magnitude of the stress is sufficient to place some fraction of fibers into the non-linear regime of their force-extension curves, the stiffness of individual fibers with respect to perturbations about this prestressed state will depend on their orientation, as schematically represented in Fig.~\ref{f:schematic}. The material response will therefore become anisotropic. Typically this prestress will also induce fiber rotation, but there is an important class of fiber networks for which this induced geometrical anisotropy can be argued to be small. Many protein fibers have been found to be well described by the wormlike chain model, in which changes to the equilibrium end-to-end separation of fiber nodes induces an entropic restoring force~\cite{Gittes1998}. For physiologically relevant parameters, such networks have been shown to strongly strain stiffen for modest strains of around 5-20\%~\cite{Storm2005}. The geometrical anisotropy will thus remain small, even though the mechanical anisotropy is significant. This corresponds to geometrically linear elasticity with a non-linear constitutive equation, also known as {\em hypoelastic} elasticity (as opposed to {\em hyperelastic} elasticity where in addition the strains are finite)~\cite{BowerBook}.

The purpose of this article is to describe analytical and numerical calculations that quantitatively predict the mechanical response of fiber networks that have become anisotropic due to an axisymmetric prestress or prestrain. Changes to network geometry are entirely neglected, allowing the consequences of this form of anisotropy to be highlighted, while still generating results of relevance to many fiber networks. The primary assumptions are hypoelasticity as described above, and also {\em affinity}, {\em i.e.} the strain field on fiber length scales is just a scaled-down version of the corresponding macroscopic strain. This assumption (the validity of which is discussed below) allows us to easily bridge the discrete and continuum representations. We also assume {\em quasi-staticity}, {\em i.e.} all calculations correspond to the elastic plateau regime of the network in question~\cite{Gittes1998,Morse1998}.

Two key results are presented. Firstly, response functions relating the displacement caused by a locally applied force are derived and reduced to a form that can be quickly integrated numerically, given network parameters and a prestress or prestrain. This can be employed in {\em e.g.} active microrheology experiments, where a probe particle is perturbed using an optical or magnetic trap~\cite{Mizuno2009}, allowing unknown parameters to be extracted {\em via} curve fitting to data. Secondly, the spatial stress field due to a force monopole is derived using a numerical procedure, and this is combined with the first result to derive predictions for the mechanical anisotropy induced by the force at a location directly in front or behind. This predictive procedure can again be employed to fit microrheology data, and is also relevant to networks with naturally-occurring force generators present, such as molecular motors in some fiber protein networks.

This paper is arranged as follows. In Sec.~\ref{s:model} the key analytical results and numerical procedures are described in detail. The basic equations are presented in Sec.~\ref{s:model1}, along with an enumeration of the reduction of independent elastic coefficients due to the specific microscopic picture assumed. Equations for the local response due to a point force are derived in Sec.~\ref{s:axisym} for axisymmetric prestrain or prestress. The numerical procedure for deriving the spatial stress field in response to a localised force monopole is explained in Sec.~\ref{s:monoThy}, and the expected domain of applicability of our assumptions are discussed in Sec.~\ref{s:thyValidity}. Applications are presented in Sec.~\ref{s:results}, where the wormlike chain model has been used throughout, allowing results to be presented in normalised forms common to all fibers that obey this model. The elastic coefficients as a function of anisotropy and magnitude of either prestrain or prestress are presented in Sec.~\ref{s:res1}, and the corresponding local response functions described in Sec.~\ref{s:res2}. The mechanical anisotropy induced by a localised force is given in Sec.~\ref{s:res3}. Finally, the relevance of our results to real networks and future directions are discussed in Sec.~\ref{s:disc}.

%
%

\section{Theory and numerical method}
\label{s:model}

In this section, analytical results for the mechanics of axisymmetric fiber networks are derived, including the local response to a point force. The numerical procedure for deriving the non-local stress field due to a point force is also explained. The notation used throughout is as follows. The line density (length of fibers per unit volume) is denoted $\rho$, the persistence length of the fibers by $\ell_{p}$, and the distance between crosslinks by $\ell_{c}$. Where relevant, fiber orientation is denoted by the unit vector~$\hat{\bf n}$.

\subsection{Anisotropic linear network elasticity}
\label{s:model1}

The linear constitutive equations for fiber networks undergoing affine deformation are now well established~\cite{Gittes1998,Morozov2011}. The assumptions here are that the unstrained network is isotropic, {\em i.e.} there is no net fiber orientation~$\hat{\bf n}$, and that the strain $u_{ij}$ remains small throughout. The stress $\sigma_{ij}$ is then given by
\begin{equation}
\sigma_{ij} = \rho \big\langle \hat{n}_{i}\hat{n}_{j}\tau(\ell_{c}\hat{n}_{k}\hat{n}_{l}u_{kl}) \big\rangle_{\hat{\bf n}}
\:.
\label{e:sigma_ij}
\end{equation}
Here $\delta\ell_{c}\equiv\ell_{c}\hat{n}_{k}\hat{n}_{l}u_{kl}$ is the extension of a fiber segment of length $\ell_{c}$ and orientation $\hat{\bf n}$ in the small-strain limit, $\tau(\delta\ell_{c})$ is the corresponding tension, and $\langle\cdots\rangle_{\hat{\bf n}}$ denotes averaging over all orientations~$\hat{\bf n}$. (\ref{e:sigma_ij}) can be derived by {\em e.g.} considering the total vector force ${\bf F}$ across an arbitrary plane with unit normal $\hat{\bf s}$, and using the definition of the stress tensor $\sigma_{ij}\hat{s}_{j}=F_{i}$.

The differential response to small changes in the strain $u_{ij}+\delta u_{ij}$ follows from expanding (\ref{e:sigma_ij}) to linear order in $\delta u_{ij}$, giving a change in stress
\begin{equation}
\delta \sigma_{ij}
=
\ell_{c}\rho
\big\langle
\tau^{\prime}(\delta\ell_{c}) \hat{n}_{i} \hat{n}_{j} \hat{n}_{k} \hat{n}_{l}
\big\rangle_{\hat{\bf n}}
\delta u_{kl}
\equiv
C_{ijkl}\delta u_{kl}\:,
\label{e:delta_sigma_ij}
\end{equation}
which defines the fourth-rank elastic stiffness tensor~$C_{ijkl}$. Contributions due to fiber rotation are assumed to be sub-dominant and are ignored. In (\ref{e:delta_sigma_ij}), $\tau^{\prime}(\delta\ell_{c})$ denotes the gradient of the force-extension curve at the prestrained fiber extension~$\delta\ell_{c}$, which depends on its orientation. If there is no prestrain, {\em i.e.} $u_{ij}\equiv0$, it is customary to drop the $\delta$'s from (\ref{e:delta_sigma_ij}), giving a more familiar expression relating $\sigma_{ij}$ to $u_{ij}$. However, (\ref{e:delta_sigma_ij}) still holds for perturbations about pre-stressed states $u_{ij}\neq0$ {\em if the assumption of small strains remains valid}, even though the force-extension relation $\tau(\delta\ell_{c})$ may become non-linear. This corresponds to the {\em hypoelastic} case described in the introduction. This limits of this and other model assumptions are discussed in Sec.~\ref{s:thyValidity}. Note that (\ref{e:delta_sigma_ij}) predicts unphysical negative moduli for force-extension curves with regions of negative slope, but such relations were not employed here.

By its definition in~(\ref{e:delta_sigma_ij}), the stiffness tensor $C_{ijkl}$ is symmetric in all perturbations of its indices.  This automatically includes all of the symmetries required of any elastic stiffness tensor, {\em i.e.} $C_{ijkl}=C_{jikl}=C_{klji}=C_{klij}$~\cite{LandauLifshitz,KachanovBook}, but also additional symmetries arising from the form of (\ref{e:delta_sigma_ij}), in which only the longitudinal filament response is present. If other modes such as bending and orientation were relevant and included to give a more complex stress tensor~\cite{Morse1998}, these symmetries may be broken. Here, however, $C_{ijkl}$ has 15 independent coefficients, which we can choose to be (in Cartesian coordinates) $C_{xxxx}$, $C_{yyyy}$, $C_{zzzz}$, $C_{xxxy}$, $C_{xxxz}$, $C_{xxyy}$, $C_{xxzz}$, $C_{xxyz}$, $C_{xyyz}$, $C_{xyzz}$, $C_{xyyy}$, $C_{xzzz}$, $C_{yyyz}$, $C_{yyzz}$ and $C_{yzzz}$. This is a reduction from the 21 independent coefficients admitted by an anisotropic elastic body in general~\cite{LandauLifshitz,KachanovBook}, which is due to the specific microscopic picture employed in deriving~(\ref{e:delta_sigma_ij}). A similar reduction arises for axisymmetric systems as discussed in Sec.~\ref{s:axisym}. In fully isotropic systems, where the Poisson ratio is fixed at $1/4$~\cite{Head2010,Ball1991}, there is only one independent coefficient ({\em i.e.} the shear modulus) rather than the usual two. It should be stressed this assumes affinity, without which (\ref{e:delta_sigma_ij}) is not generally valid and the number of independent moduli may increase.

\subsection{Local mechanical response with axisymmetry}
\label{s:axisym}

We now turn to consider axisymmetric (also known as transversely isotropic~\cite{KachanovBook}) systems that are rotationally symmetric about a fixed axis. Without loss of generality, Cartesian coordinates $(x,y,z)$ are used with the $z$-axis taken to be the axis of symmetry. The elastic stiffness tensor $C_{ijkl}$ now has fewer independent coefficients than the generally anisotropic case: $C_{ijkl}$ is invariant under changes $x\leftrightarrow y$, $C_{ijkl}\equiv0$ if there are an odd number of $x$'s or $y$'s in the indices, and $C_{xxxx}=3C_{xxyy}$ as seen from performing the part of the integral (\ref{e:delta_sigma_ij}) around the $z$-axis. Taking these additional constraints into account, there remain 3 independent elastic coefficients, here taken to be $C_{xxxx}\equiv C_{0}$, $C_{xxzz}\equiv C_{2}$ and $C_{zzzz}\equiv C_{4}$. This is a reduction from the 5 independent moduli for elastic bodies with the same symmetry~\cite{KachanovBook}, again reflecting the specific microscopic picture assumed. A similar reduction (of 5 to 3) occurs for nematic liquid crystals but for different reasons~\cite{deGennesProst}.

To determine the local mechanical response, a virtual point force ${\bf f}^{\rm virt}$ is applied at the origin, giving the force balance equations~\cite{LandauLifshitz},
\begin{equation}
\partial_{j}\delta\sigma_{ij}+f_{i}^{\rm virt}\delta({\bf x})=0\:,
\label{e:fvirt}
\end{equation}
where the delta function localises the force at the origin. Here and below, $\delta\sigma_{ij}$ and $\delta u_{ij}$ denote small changes to the stress and strain tensor about their prestressed or prestrained values $\sigma_{ij}$ and $u_{ij}$ respectively. The $\partial_{j}\delta\sigma_{ij}$ in (\ref{e:fvirt}) can be written in terms of the first derivative of the strain fluctuations $\delta u_{ij}$ using~(\ref{e:delta_sigma_ij}) (assuming constant $C_{ijkl}$), and hence the second-derivatives of the displacement fluctuations $\delta u_{i}$ using standard formulae~\cite{LandauLifshitz}. This is then Fourier transformed as per $\delta\tilde{u}_{i}({\bf q})\equiv\int e^{-i{\bf q}\cdot{\bf x}}\delta u_{i}({\bf x}){\rm d}{\bf x}$ to give the matrix equation
\begin{equation}
{\bf f}^{\rm virt}
=
M
\delta\tilde{\bf u}
\label{e:matrix}
\end{equation}
where using the notation ${\bf q}=(q_{x},q_{y},q_{z})$,
\begin{widetext}
\begin{equation}
M
=
\left(
\begin{array}{ccc}
C_{0}q^{2}_{x}+\frac{1}{3}C_{0}q^{2}_{y}+C_{2}q^{2}_{z} & \frac{2}{3}C_{0}q_{x}q_{y} & 2C_{2}q_{x}q_{z}
\\
\frac{2}{3}C_{0}q_{x}q_{y} & \frac{1}{3}C_{0}q_{x}^{2} + C_{0}q_{y}^{2} + C_{2}q_{z}^{2} & 2C_{2}q_{y}q_{z}
\\
2C_{2}q_{x}q_{z} & 2C_{2}q_{y}q_{z} & C_{2}(q_{x}^{2}+q_{y}^{2}) + C_{4}q_{z}^{2}
\end{array}
\right)\:.
\nonumber
\end{equation}
\end{widetext}

In order to obtain the spatial response field due to ${\bf f}^{\rm virt}$, $M$ is inverted and inserted into (\ref{e:matrix}), which would properly then be inverse Fourier transformed to give $\delta{\bf u}({\bf x})$; however this inverse transform is not straightforward to perform. Instead we restrict attention to the displacement ${\bf u}^{\rm probe}=(u_{x}^{\rm probe},u_{y}^{\rm probe},u_{z}^{\rm probe})$ of a sphere of radius $a$ at the origin to which the force ${\bf f}^{\rm virt}$ is applied, which allows an approximate solution to be readily attained~\cite{Alex2000}. In this procedure, the $q$-modes are truncated at a maximum magnitude $|q|^{\rm max}=\pi/2a$, and the reverse Fourier transform performed with ${\bf x}=0$. This reverse transform cannot be performed analytically for arbitrary $C_{0}$, $C_{2}$ and $C_{4}$, but can be reduced by standard techniques to two one-dimensional integrals that can be easily performed numerically,
\begin{widetext}
\begin{eqnarray}
u_{x}^{\rm probe}
&=&
\frac{f_{x}^{\rm ext}}{4\pi a}
\int^{1}_{0}{\rm d}s\,
\frac
{\frac{2}{3}C_{0}C_{2}
+\left[\frac{2}{3}C_{0}C_{4}-\frac{4}{3}C_{0}C_{2}-C_{2}^{2}\right]s^{2}
+\left[C_{2}C_{4}-\frac{2}{3}C_{0}C_{4}+C_{2}^{2}+\frac{2}{3}C_{0}C_{2}\right]s^{4}}
{A(s)B(s)}
\:,
\nonumber
\\
u_{z}^{\rm probe}
&=&
\frac{f_{z}^{\rm ext}}{12\pi a}
\int^{1}_{0}{\rm d}s\,
\frac{C_{0}+(C_{2}-C_{0})u^{2}}{B(s)}
\:,
\label{e:uprobe}
\end{eqnarray}
\end{widetext}
where
\begin{eqnarray}
A(s)
&=&
C_{0}+(3C_{2}-C_{0})s^{2}\:,
\nonumber\\
B(s) &= &
\frac{1}{3}C_{0}C_{2}
+
\left[\frac{1}{3}C_{0}C_{4}-C_{2}^{2}-\frac{2}{3}C_{0}C_{2}\right]s^{2}
\nonumber\\
&&+
\left[\frac{1}{3}(C_{2}C_{4}-C_{0}C_{4}+C_{0}C_{2})+C_{2}^{2}\right]s^{4}
\:.
\end{eqnarray}
The $u^{\rm probe}_{y}$ equation is similar to that of $u_{x}^{\rm probe}$, with $f_{x}^{\rm virt}$ replaced by $f_{y}^{\rm virt}$. It is conventional to express results in terms of the response functions in directions parallel and perpendicular to the axis of material symmetry,
\begin{eqnarray}
\alpha^{\parallel}&=&u_{z}^{\rm probe}/f^{\rm virt}_{z}\:,\nonumber\\
\alpha^{\perp}&=&u_{x}^{\rm probe}/f_{x}^{\rm virt}\:.
\label{e:alphas}
\end{eqnarray}
For an isotropic material, $C_{0}=3C_{2}=C_{4}=3G$ with $G$ the shear modulus, and (\ref{e:uprobe}) and (\ref{e:alphas}) can be evaluated analytically,
\begin{equation}
\alpha^{\parallel}=\alpha^{\perp}=\frac{7}{36\pi aG}\:,
\label{e:alpha0}
\end{equation}
matching real-space calculations~\cite{Alex2001} for an isotropic elastic body with the Poisson ratio of $\frac{1}{4}$ expected for affinely-deforming networks in 3D~\cite{Head2010,Ball1991}.

\subsection{Anisotropy and non-linearity induced by a force monopole}
\label{s:monoThy}

Sec.~\ref{s:axisym} explains how to calculate the response given the differential stiffness tensor~$C_{ijkl}$, which in turn depends on the prestress $\sigma_{ij}$ or prestrain~$u_{ij}$. Here we explain how to calculate numerically $\sigma_{ij}$ and $u_{ij}$ for a force monopole applied to an initially isotropic network. If the material is deformed at any given point that obeys axisymmetry, and the fiber force-extension relation $\tau(\delta\ell_{c})$ is known, then $C_{0}$, $C_{2}$ and $C_{4}$ can be evaluated using (\ref{e:delta_sigma_ij}), and inserted into (\ref{e:uprobe}) to determine the local response functions $\alpha^{\parallel}$, $\alpha^{\perp}$ (\ref{e:alphas}) at that point. This procedure is here applied to determine both response functions at various distances in front of and behind an external force monopole ${\bf f}^{\rm ext}$ along the axis of symmetry, for which, assuming there is no spontaneous symmetry breaking induced by {\em e.g.} elastic instabilities, axisymmetry will hold. Even assuming hypoelasticity, it is difficult to solve the full problem analytically if $\tau(\delta\ell_{c})$ is non-linear, hence the spatial response due to ${\bf f}^{\rm ext}$ is here determined numerically.


To perform the numerical calculations, cylindrical coordinates $(r,\theta,z)$ were employed with the $z$-axis aligned parallel to the direction of the applied force ${\bf f}^{\rm ext}$, and variation with $\theta$ suppressed following the expected axial symmetry of the solution. Displacement vectors were defined at regular $(r,z)$ lattice nodes and the corresponding strains $u_{ij}$ were interpolated onto an interpenetrating staggered mesh by first-order finite differencing. The local strains were then converted to changes in fiber segment length $\delta\ell_{c}$ by assuming affine deformation, {\em i.e.} $\delta\ell_{c}=\ell_{c}\hat{n}_{i}\hat{n}_{j}u_{ij}$ for an orientation~$\hat{\bf n}$. This was then converted to a longitudinal tension $\tau(\delta\ell)$ using the chosen single fiber model. For this calculation the extensible wormlike chain model was used, which generalises the inextensible model by the inclusion of a contour modulus~$K$~\cite{Storm2005},
\begin{equation}
\delta\ell_{c} = \frac{\ell_{c}^{2}}{\ell_{p}}\left(1+\frac{\tau}{K}\right)
u^{\rm inext}\left(\frac{\tau\ell_{c}^{2}}{\kappa\pi^{2}}\left[1+\frac{\tau}{K}\right]\right)
+\frac{\tau}{K}\ell_{0}\:,
\label{e:wlc1}
\end{equation}
where $u^{\rm inext}(\phi)$ is the dimensionless inextensible wormlike chain expression
\begin{equation}
u^{\rm inext}(\phi)
=
\frac{1}{6}
-
\frac{1}{2\pi^{2}\phi}\left[
\pi\sqrt{\phi}\coth(\pi\sqrt{\phi})-1
\right]\:.
\label{e:wlc2}
\end{equation}
$\ell_{0}=\ell_{c}-\ell_{c}^{2}/6\ell_{p}$ is the natural end-to-end distance and $\kappa=k_{\rm B}T\ell_{p}$ the bending rigidity. The inextensible model is recovered in the limit $K/\tau\rightarrow\infty$, {\em i.e.}
\begin{equation}
\delta\ell_{c}=\frac{\ell_{c}^{2}}{\ell_{p}}u^{\rm inext}\left(\frac{\tau\ell_{c}^{2}}{\kappa\pi^{2}}\right)\:.
\label{e:wlc3}
\end{equation}
For $\phi<-1$, (\ref{e:wlc2}) is undefined and we assume such fibers have buckled and no longer contribute to the mechanical response of the network. We employed a high contour stiffness, $K$=500pN for vimentin-like network parameters and $K$=3000pN for actin-like networks, but varying $K$ made little difference to the predictions and was primarily employed to remove the singularity at $u=\frac{1}{6}$ in the (inextensible) force-extension curve, aiding numerical convergence.

Numerically inverting (\ref{e:wlc1}) and (\ref{e:wlc2}) gives the tension $\tau$ as a function of extension $\delta\ell_{c}$ and hence strain $u_{ij}$, which is then integrated over all orientations to give the stress field $\sigma_{ij}$ using~(\ref{e:sigma_ij}). The equations of mechanical equilibrium in cylindrical coordinates with $\theta$-variation suppressed are~\cite{KachanovBook}
\begin{eqnarray}
0 & = & \partial_{r}\sigma_{rr}+\partial_{z}\sigma_{rz}+\frac{\sigma_{rr}-\sigma_{\theta\theta}}{r}+f^{\rm ext}_{r}W({\bf x})\:,
\nonumber\\
0 & = & \partial_{r}\sigma_{rz} + \partial_{z}\sigma_{zz} + \frac{\sigma_{rz}}{r} + f^{\rm ext}_{z}W({\bf x})
\label{e:eqm}
\end{eqnarray}
at each node, and our goal is to find the displacement field corresponding to global equilibrium. Here the external force is localised at the origin by the function $W({\bf x})$, which integrates to one and decays to zero for large $|{\bf x}|$. However, we do {\em not} use a delta function as this introduces sharp gradients and numerical difficulties. Instead a smooth Gaussian profile was used, $W({\bf x})=\frac{1}{\sqrt{2\pi\sigma^{2}}}e^{-(r^{2}+z^{2})/2\sigma^{2}}$, with $\sigma=1\mu$m.

To solve (\ref{e:eqm}), Newton's method was used as follows. The displacements at each interior node was written as a combined vector ${\bf U}$, where ${\bf U}$ contains the radial and axial components of each nodal displacement. The internal forces $\partial_{j}\sigma_{ij}$ were rewritten as a nodal force vector ${\bf F}^{\rm int}$ arranged identically to ${\bf U}$. Matching ${\bf F}^{\rm int}$ was a constant vector ${\bf F}^{\rm ext}$ giving the external force (combined with $W({\bf x})$) for each node. Each component of ${\bf F}^{\rm int}$ was determined from the relative displacements of nearby nodes, and was therefore a smooth function of ${\bf U}$ that can be expanded about a given ${\bf U}_{0}$, giving the global equilibrium equation
\begin{equation}
0
=
F^{\rm ext}_{\alpha}  + F^{\rm int}_{\alpha}({\bf U})
=
F^{\rm ext}_{\alpha} +  F^{\rm int}_{\alpha}({\bf U}_{0}) + A_{\alpha\beta}\delta U_{\beta} + {\mathcal O}(\delta U^{2})
\label{e:eqm2}
\end{equation}
where ${\bf U}={\bf U}_{0}+\delta{\bf U}$ and the $\alpha$, $\beta$ subscripts refer to coordinates in the ${\bf U}$ and ${\bf F}$ vectors. Here, $A_{\alpha\beta}=\partial F^{\rm int}_{\beta}/\partial U_{\alpha}$ is a large, symmetric stiffness matrix. Neglecting the quadratic and higher terms in $\delta{\bf U}$ in (\ref{e:eqm2}) gives a linear equation that can be solved {\em via} matrix inversion,
\begin{equation}
\delta U_{\alpha}=-A^{-1}_{\alpha\beta}\left[F^{\rm ext}_{\beta}+F^{\rm int}_{\beta}({\bf U}_{0})\right]\:.
\label{e:inv}
\end{equation}
The full non-linear equations are then re-linearised about this new point ${\bf U}_{1}={\bf U}_{0}+\delta{\bf U}$ and inverted once more, producing a succession of estimates ${\bf U}_2$, ${\bf U}_{3}$ {\em etc.} that obey equilibrium in their corresponding linearised systems. Iteration continues until the largest change in any single component of ${\bf U}$ changes by less than some small threshold value, which was taken to be $10^{-5}\mu$m here.

Dirichlet boundary conditions corresponding to the known linear solution~\cite{LandauLifshitz} for a point force ${\bf f}^{\rm ext}$ were imposed. Varying the linear system size by a factor of 2 had only slight effects (roughly 1\%) on the stress field, so our findings are not sensitive to this choice of boundary condition. Results presented here correspond to $0\leq r\leq80\mu m$ and $-80\mu{\rm m}\leq z\leq 80\mu{\rm m}$, with a mesh size of 0.4$\mu$m in both directions. The matrix inversion (\ref{e:inv}) was solved using the conjugate gradient method~\cite{NumericalRecipes}.

\subsection{Domain of validity}
\label{s:thyValidity}

Here we clarify the assumptions underlying our modelling approach and their expected domain of validity. Firstly, it has been assumed throughout that the network geometry without prestrain is isotropic, in keeping with our aim of quantifying, and hence elucidating, the effects of anisotropy induced by internal or external perturbations alone. It will not immediately apply to networks that are anisotropic in their unstressed state.

It is possible to be precise regarding the {\em hypoelastic} limit assumed in Sec.~\ref{s:model1}, {\em i.e.} adopting the geometrically linear (small strain) limit while allowing the mechanical properties to become non-linear. The characteristic stress at which networks of wormlike chains exhibit non-linearities has been estimated as $\sigma_{c}=\rho k_{\rm B}T\ell_{p}/\ell_{c}^{2}$~\cite{Lin2010}. Coupled with the linear shear modulus $G^{\rm lin}=6\rho k_{\rm B}T\ell_{p}^{2}/\ell_{c}^{3}$, this suggests the strain at which mechanical non-linearities arise is approximately
\begin{equation}
\gamma_{c}
\equiv
\frac{\sigma_{c}}{G^{\rm lin}}
=
\frac{\ell_{c}}{6\ell_{p}}\:.
\label{e:gamma_c}
\end{equation}
The hypoelastic assumption is applicable to strains $\gamma$ obeying $\gamma_{c}\lesssim\gamma\ll1$. For typical actin networks, $\ell_{c}\approx1-3\mu$m and $\ell_{p}\approx17\mu$m, suggesting $\gamma_{c}$ is of only a few percent, consistent with experiments~\cite{Storm2005} and confirming validity of the hypoelastic assumption for a broad range of strains. For vimentin~\cite{Lin2010}, $\gamma_{c}$ is larger, around 10$\%$, giving a reduced, but still finite, range of validity.

Conservative limits on the validity of the affine assumption come from two sources. Firstly, theoretical and experimental studies suggest uniform strains are affine when the filament length $L$ exceeds an intrinsic length $\lambda=\ell_{c}\sqrt[3]{\ell_{c}/\ell_{b}}$ with $\ell_{b}$ a length related to the filament diameter~\cite{Head2003a,Liu2007}. This is certainly realisable in actin and vimentin systems~\cite{Gardel2004,Lin2010}. However, when strain gradients are present, affinity is not expected to apply when the strain significantly varies between crosslinks. This situation is discussed (with quantitative estimates) in Sec.~\ref{s:monoThy}, where it is estimated that affinity would break down within 1$\mu$m of a 10pN force for typical actin networks. However, this estimate is likely too conservative as it ignores strain stiffening. Furthermore, recent experiments on vimentin networks have shown how the range of validity of these calculations can be extended to give quantitative agreement with the data, as explained in detail elsewhere~\cite{HeadSub}. Further investigation is required to rigorously delineate the limits of these predictions, possibly {\em via} microscopic numerical modelling.

%
%
\section{Applications}
\label{s:results}

Here we present predictions of the calculations of Sec.~\ref{s:model} for fibers obeying the wormlike chain model. This model was chosen since the only alternative in common usage, {\em i.e.} elastic beam models~\cite{Head2003a,Head2003b,Missel2010,Ball1991,Head2005}, exhibit no non-linearity in their force-extension curves and hence no anisotropy in the hypoelastic limit. The moduli and response functions for a predefined prestress is considered first, before combining these calculations with the spatial response field generated by a  force monopole.

\subsection{Moduli for a given prestress or prestrain}
\label{s:res1}

\begin{figure}[htbp]
\centerline{\includegraphics[width=8cm]{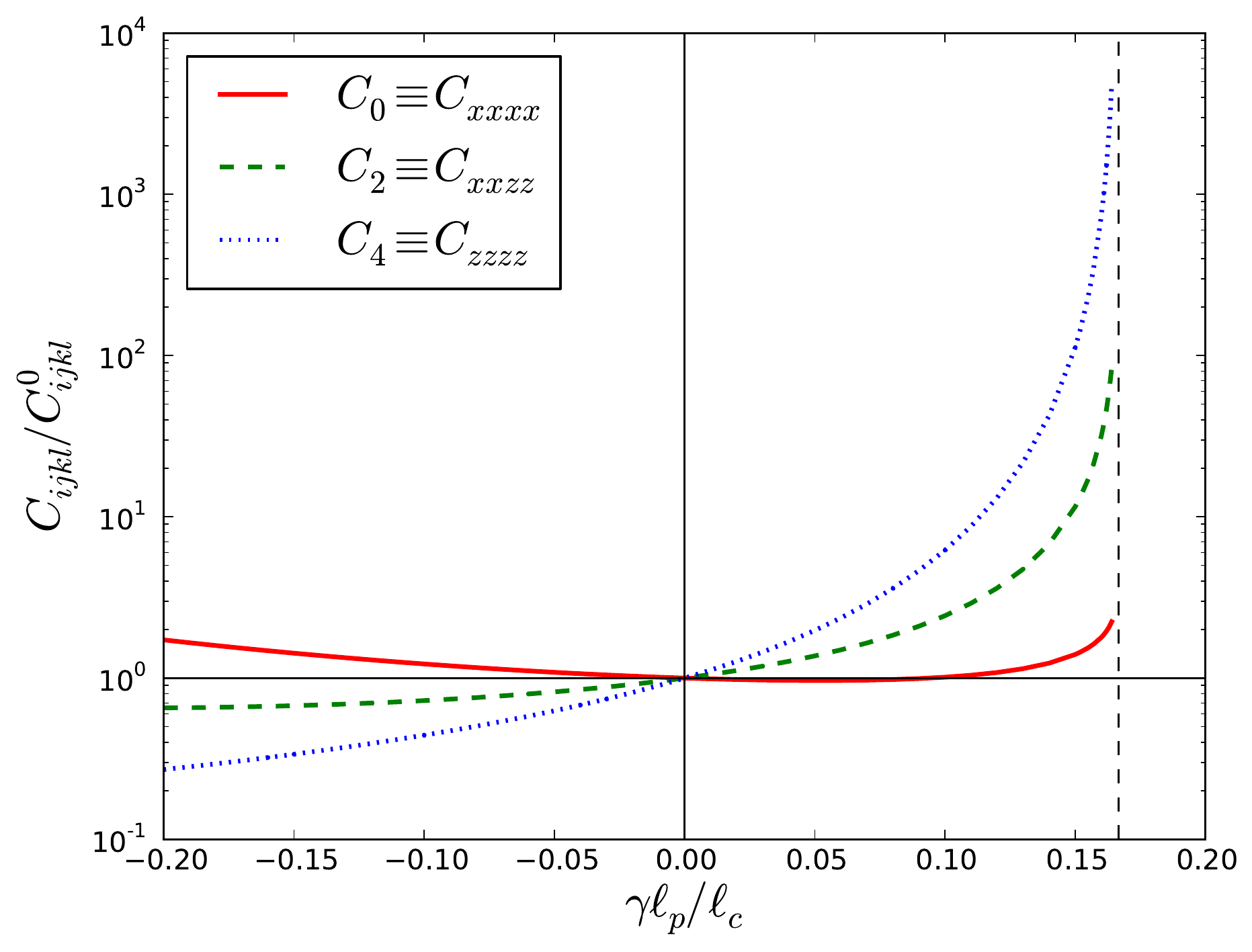}}
\caption{The independent elastic moduli $C_{0}$, $C_{2}$ and $C_{4}$ as a function of the normalised prestrain magnitude~$\gamma\ell_{p}/\ell_{c}$, for a Poisson ratio $\nu=\frac{1}{4}$. Each modulus has been scaled to its zero prestrain values~$C_{ijkl}^{0}$, which obey $C^{0}_{0}=3C^{0}_{2}=C^{0}_{4}$. The vertical dashed line is at $\gamma\ell_{p}/\ell_{c}=\frac{1}{6}$. The thin horizontal and vertical solid lines are to guide the eye to the $\gamma=0$ solution.}
\label{f:AsPreStrain}
\end{figure}

Although in many applications the perturbation will be an applied force, thus generating a prestress, it is also insightful to consider prestrains. For an axisymmetric prestrain with no transverse normal stresses, the strain tensor can be written (in Cartesian coordinates, with the axis of symmetry along the $z$-axis) as
\begin{equation}
u_{ij}=\gamma\,{\rm diag}\left(-\nu,-\nu,1\right)\:,
\end{equation}
where the dimensionless $\nu$ parameterises the anisotropy (for linear response, $\nu$ is simply the Poisson ratio). Assuming affinity, and taking $\nu=\frac{1}{4}$~\cite{Head2010,Ball1991}), filaments aligned at an angle $\theta$ to the axis of symmetry will be extended by a relative amount
\begin{equation}
\frac{\delta\ell_{c}}{\ell_{c}}=\gamma
\left(
\cos^{2}\theta
-
\frac{1}{4}\sin^{2}\theta
\right)\:.
\label{e:prestrain}
\end{equation}
This can then be inserted into (\ref{e:delta_sigma_ij}), along with a specific choice of $\tau(\ell_{c})$, to determine the elastic moduli. For the wormlike chain model, the tensions and extensions can be expressed in normalised forms by scaling the prefactor to $\gamma\ell_{p}/\ell_{c}=(\delta\ell_{c}/\ell_{c})(\ell_{p}/\ell_{c})$, which then fully specifies the magnitude of the prestrain as per~(\ref{e:wlc3}). Fig.~\ref{f:AsPreStrain} shows the independent elastic coefficients for varying $\gamma$ with fixed $\nu=1/4$, showing the expected stiffening along the $z$-axis with $\gamma>0$, with $C_{4}>C_{2}>C_{0}$. For $\gamma<0$, the order becomes $C_{0}>C_{2}>C_{4}$, and transverse modes now become stiffer due to their extension (for $\nu>0$). Also note the divergence at $\gamma\ell_{p}/\ell_{c}=\frac{1}{6}$ which arises when the end-to-end distance of fiber segments aligned with the $z$-axis equals their contour length, and can extend no more without contour stretching.

For prestresses rather than prestrains, it is necessary to write down the tension of a filament as a function of its orientation, in analogy to the extension relation~(\ref{e:prestrain}). Here we follow the same method, and the same notation, as Morozov and Pismen~\cite{Morozov2011}. Two parameters are employed, a characteristic tension $\tau_{0}$, and a dimensionless quantity $\beta$ that parameterises the degree of anisotropy in the prestress. Then the tension in a fiber aligned with an angle $\theta$ to the axis of symmetry is~\cite{Morozov2011}
\begin{equation}
\tau(\theta) = \tau_{0}\left(
\cos^{2}\theta
+
\frac{1-\beta}{1+\beta}
\sin^{2}\theta
\right)\:,
\label{e:prestress}
\end{equation}
where $\tau(\theta)>0$ corresponds to tension and $\tau(\theta)<0$ to compression. Assuming $|\beta|<1$, $\beta\tau_{0}>0$ corresponds to a more positive stress in the $z$-direction, and $\beta\tau_{0}<0$ to more positive stresses in the transverse plane. The prestress is isotropic or absent if $\beta\tau_{0}=0$. Note there is no simple relationship between (\ref{e:prestrain}) and~(\ref{e:prestress}), even in linear response, as each has been chosen to conform to simple forms to facilitate interpretation of the results and comparison to previous work.

\begin{figure}[htbp]
\centerline{\includegraphics[width=9cm]{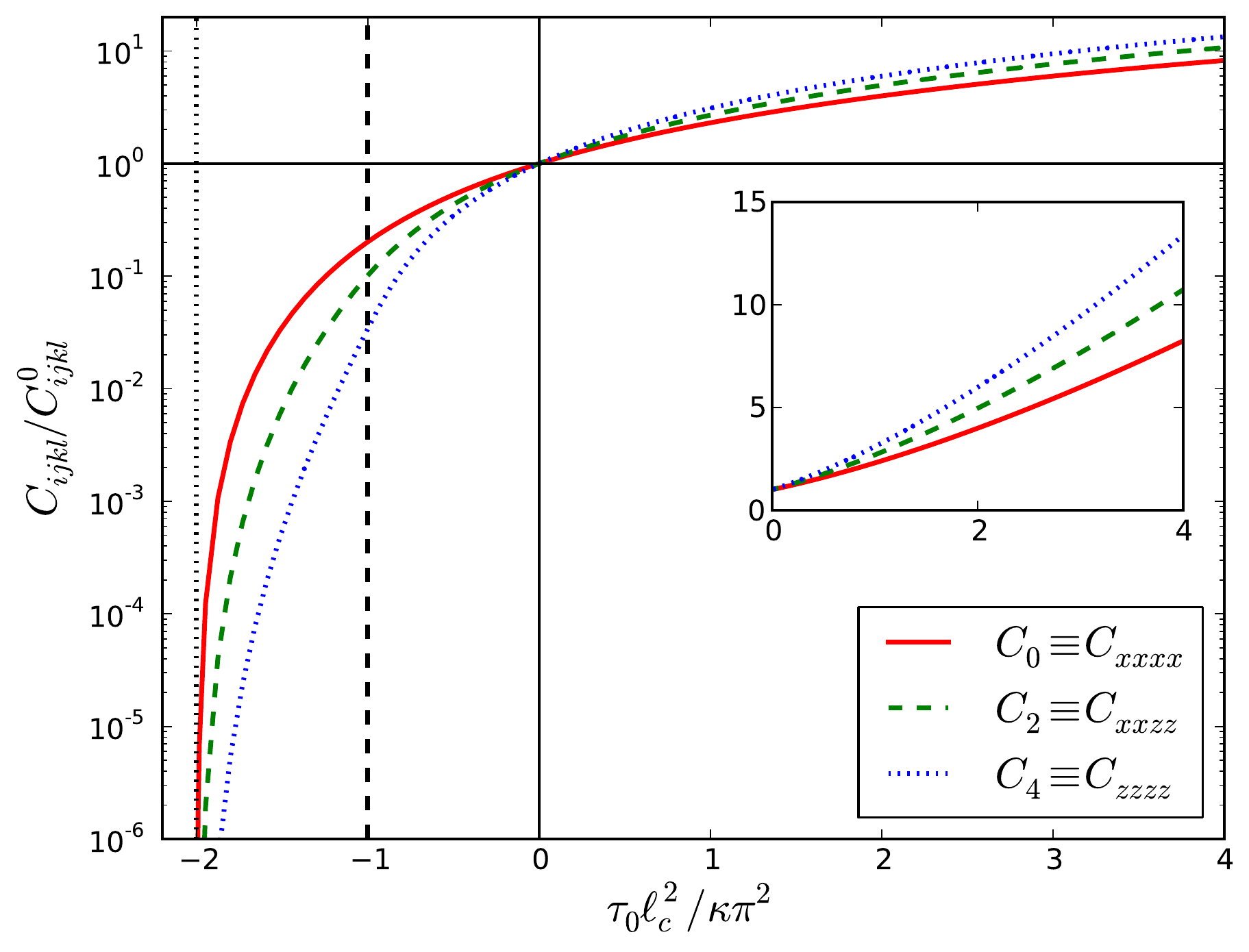}}
\caption{The independent elastic moduli for an axisymmetric prestress of normalised magnitude $\tau_{0}\ell_{c}^{2}/\kappa\pi^{2}$, for an anisotropy parameter $\beta=\frac{1}{3}$. The moduli are scaled to their zero prestress values $C_{ijkl}^{0}$. The vertical dashed and dotted lines correspond to $\tau_{0}\ell_{c}^{2}/\kappa\pi^{2}=-1$ and $-2$ respectively. The thin horizontal and vertical solid lines are to guide the eye to the $\tau_{0}=0$ solution. The inset shows the same data plotted with both axes linear.}
\label{f:As}
\end{figure}

For the wormlike chain model the prestress magnitude $\tau_{0}$ can be normalised to $\tau_{0}\ell_{c}^{2}/\kappa\pi^{2}$ as per~(\ref{e:wlc3}). Determining the elastic coefficients using (\ref{e:delta_sigma_ij}) now requires converting each tension $\tau$ to an extension, and expanding the force-extension curve about this point, which is straightforward to perform numerically. The variation of the elastic moduli with $\tau_{0}$ for a fixed anisotropy $\beta=\frac{1}{3}$ is presented in Fig.~\ref{f:As}. There is a clear stiffening of all moduli for networks under tension $\tau_{0}>0$, with $C_{4}>C_{2}>C_{0}$, and a corresponding softening for $\tau_{0}<0$ with $C_{0}>C_{2}>C_{4}$. Also marked on this figure are two special tensions: $\tau_{0}\ell_{c}^{2}/\kappa\pi^{2}=-1$, which is when fibers aligned with the $z$-axis become contracted beyond the range of the wormlike chain model, which we interpret as a buckling event beyond which the single-fiber response is identically zero. There is however no clear signature of this buckling until $\tau_{0}\ell_{c}^{2}/\kappa\pi^{2}=(1+\beta)/(1-\beta)=-2$ for this~$\beta$, when all fibers, including those oriented transversely to the $z$-axis, buckle. All linear elastic moduli equal zero at and below this point.

Also shown in the inset to Fig.~\ref{f:As} is the same data plotted on linear axes, demonstrating approximate proportionality between prestress and stiffness, which has also been observed in isolated smooth muscle cells~\cite{Park2010}. We note however that there is a slight upturn to the model predictions when crossing from linear prestress $\tau_{0}\ell_{c}^{2}/\kappa\pi^{2}\ll1$ to non-linear $\tau_{0}\ell_{c}^{2}/\kappa\pi^{2}\gg1$, which is not apparent in the experimental data. It is possible that the experimental data is all in the non-linear regime. Alternatively, one of the model assumptions may have broken down, or some relevant feature is absent. Measurements of well-controlled {\em in vitro} systems would help clarify this deviation.

\subsection{Local response in a predefined stress field}
\label{s:res2}

\begin{figure}[htbp]
\centerline{\includegraphics[width=9cm]{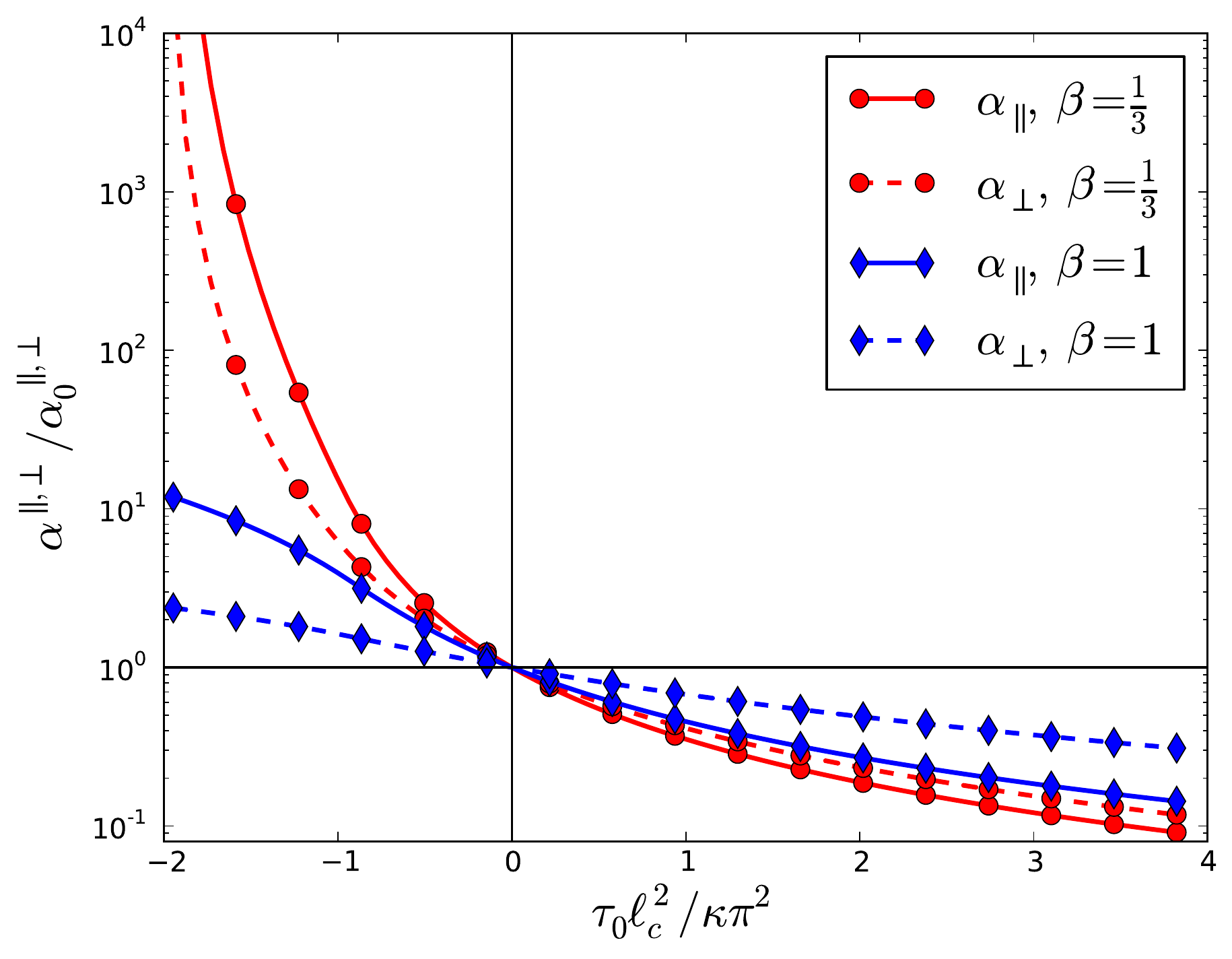}}
\caption{Response functions in parallel and perpendicular directions to the axis of symmetry versus normalised prestress $\tau_{0}$ for $\beta=\frac{1}{3}$ and $\beta=1$ as denoted in the legend. The response functions are normalised to their zero prestress $\alpha_{0}^{\parallel,\perp}$.}
\label{f:probeVsTau0}
\end{figure}

The local displacement ${\bf u}$ due to a point force ${\bf f}$ can be quantified by the response functions $\alpha^{\parallel}=u_{z}/f_{z}$ and $\alpha^{\perp}=u_{x}/f_{x}$ as described in Sec.~\ref{s:axisym}, so higher $\alpha$ correspond to a softer material. Continuing with prestresses~(\ref{e:prestress}), $\alpha^{\parallel,\perp}$ can be numerically evaluated using (\ref{e:uprobe}). Examples for $\beta=\frac{1}{3}$ and $\beta=1$ are shown in Fig.~\ref{f:probeVsTau0}. The trends observed are consistent with the observations of the previous section, {\em i.e.} a stiffening for $\tau_{0}>0$ with $\alpha^{\parallel}<\alpha^{\perp}$ and both less than their zero prestress values, with the opposite trend for $\tau_{0}<0$. The divergence of both $\alpha^{\parallel}$ and $\alpha^{\perp}$ at $\tau_{0}\ell_{c}^{2}/\kappa\pi^{2}=-2$ (for $\beta=\frac{1}{3}$) corresponds to the vanishing of the elastic coefficients in Fig.~\ref{f:As}. There is no divergence for $\beta=1$, since for this extreme anisotropy fibers oriented perpendicular to the $z$-axis never become prestressed.

\begin{figure}[htbp]
\centerline{\includegraphics[width=9cm]{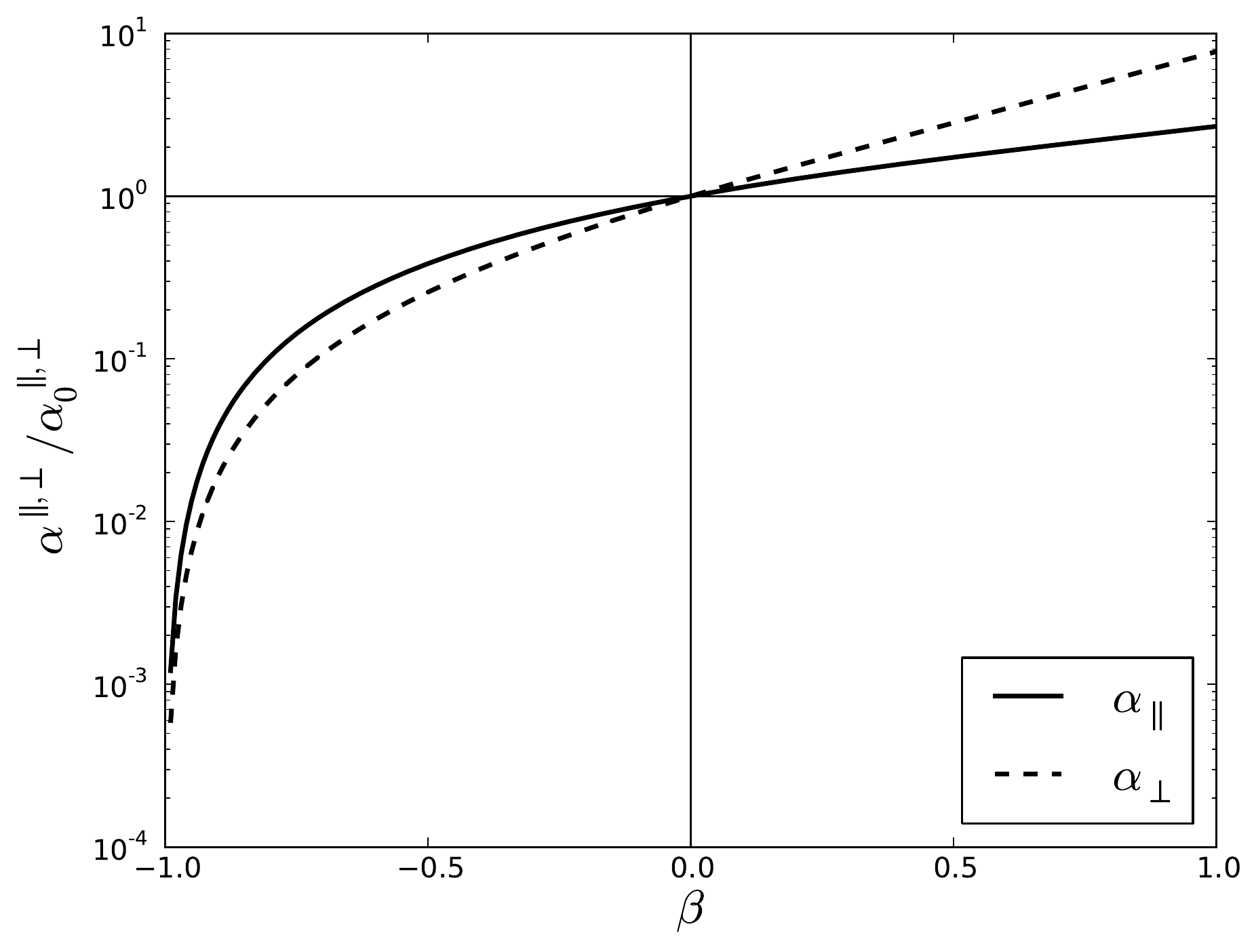}}
\caption{Response functions versus anisotropy $\beta$ for fixed $\tau_{0}\ell_{c}^{2}/\kappa\pi^{2}=1$.}
\label{f:probeVsBeta}
\end{figure}

It is evident from Fig.~\ref{f:probeVsTau0} that, for all $\tau_{0}$, the degree of stiffening or softening for $\beta=1$ is reduced compared to $\beta=\frac{1}{3}$, reflecting the lower net tension for a given $\tau_{0}$ in~(\ref{e:prestress}). This effect is more clearly seen by plotting the variation with $\beta$ for a fixed~$\tau_{0}>0$ as in Fig.~\ref{f:probeVsBeta}, demonstrating that the material is softer in both directions for $\beta>0$, but more so in transverse directions, with the opposite trend for $\beta<0$. The vanishing $\alpha^{\parallel,\perp}$ as $\beta\rightarrow-1$ corresponds to a diverging pretension for transversely-aligned filaments as per (\ref{e:prestress}).

\subsection{Anisotropy due to a force monopole}
\label{s:res3}

Localised perturbations to the network, in whatever form they take, that are of sufficient magnitude to generate a non-linear elastic response can induce anisotropy in a spatially-varying manner. For simplicity we consider here just the case of a force monopole that can be generated by {\em e.g.} perturbing a probe particle with an optical or magnetic trap, and focus attention on points directly behind and in front of the force, for which the axisymmetric theory developed here should be valid. This protocol has been recently applied to vimentin networks~\cite{HeadSub}. The methodology was outlined in Sec.~\ref{s:monoThy}.

\begin{figure}[htbp]
\centerline{\includegraphics[width=9cm]{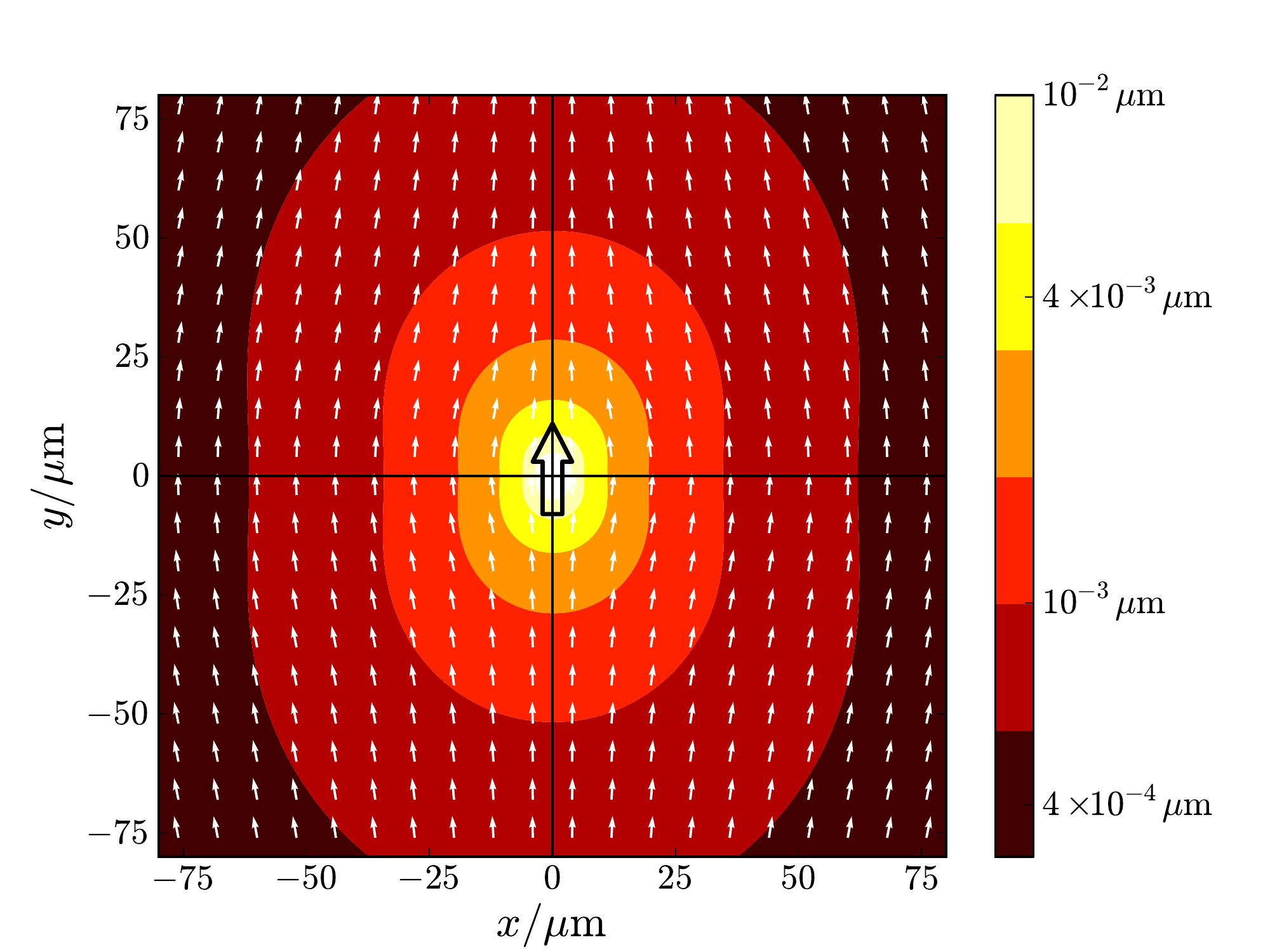}}
\caption{The displacement field for an applied force of 1pN in Euler (pre-strain) coordinates for $\ell_{p}=0.5\mu$m, $\ell_{c}=0.4\mu$m and $\rho=16.25\mu{\rm  m}^{-2}$. The direction of material displacements are given as white arrows, and the magnitude given by the contours as denoted by the calibration bar. The location and direction of the applied force is denoted by the large, open black arrow at the center.}
\label{f:disp1}
\end{figure}

\begin{figure}[htbp]
\centerline{\includegraphics[width=9cm]{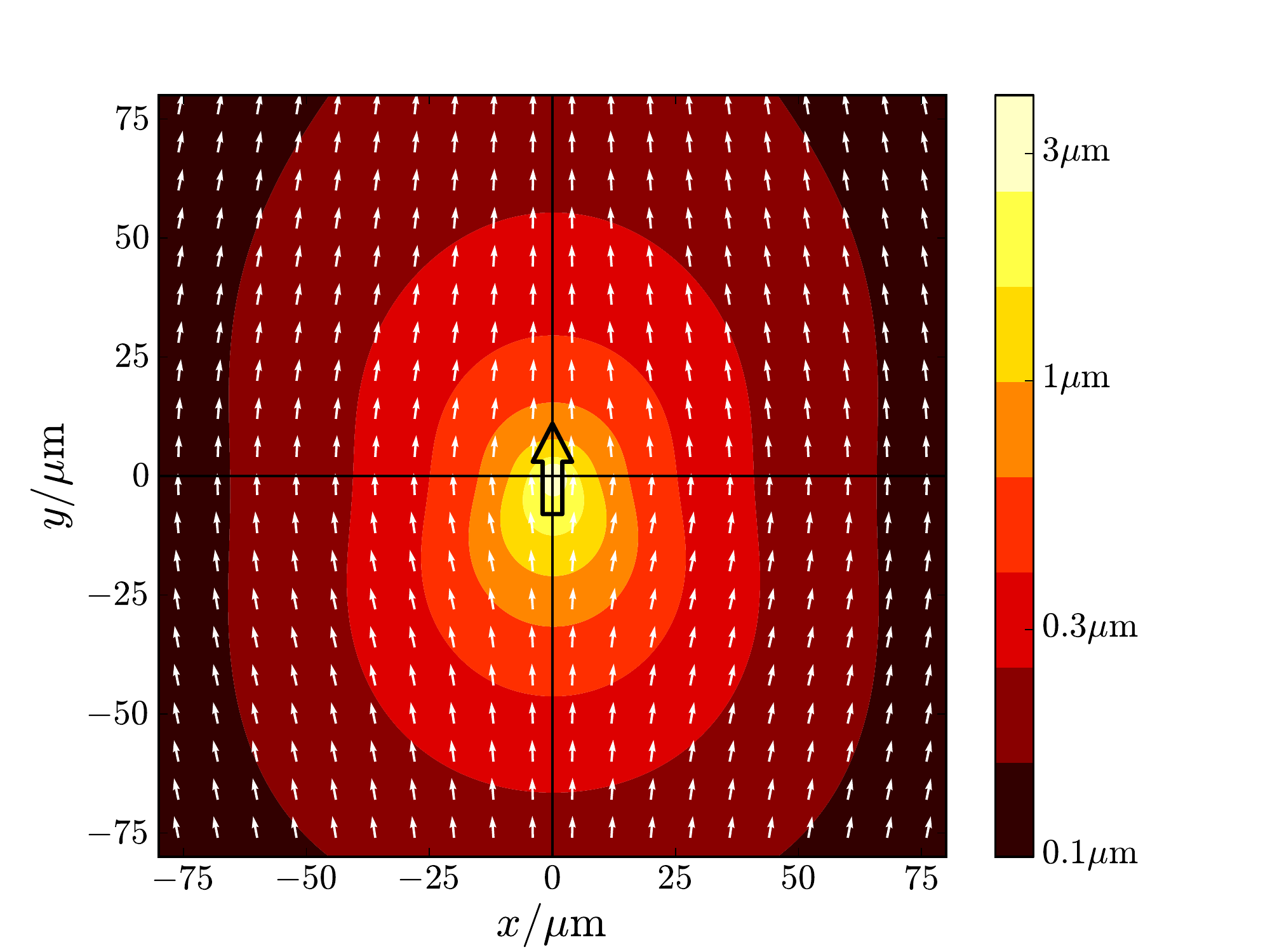}}
\caption{The same as Fig.~\ref{f:disp1} for an applied force of 300pN.}
\label{f:disp2}
\end{figure}

Examples of the displacement fields induced by a small (1pN) and large (300pN) applied forces are presented in Figs.~\ref{f:disp1} and \ref{f:disp2} respectively, for material parameters representative of the vimentin networks of~\cite{HeadSub}. For the smaller force, the displacement field is symmetric behind and in front of the force, and indeed can be shown to obey the the linear solution except near the point of force application. By contrast, for the larger force there is a marked fore-aft asymmetry in which the displacements behind the probe exceed those in front. This is a consequence of the strain stiffening under tension and softening under compression that is inherent to the wormlike chain model.

\begin{figure}[htbp]
\centerline{\includegraphics[width=8.7cm]{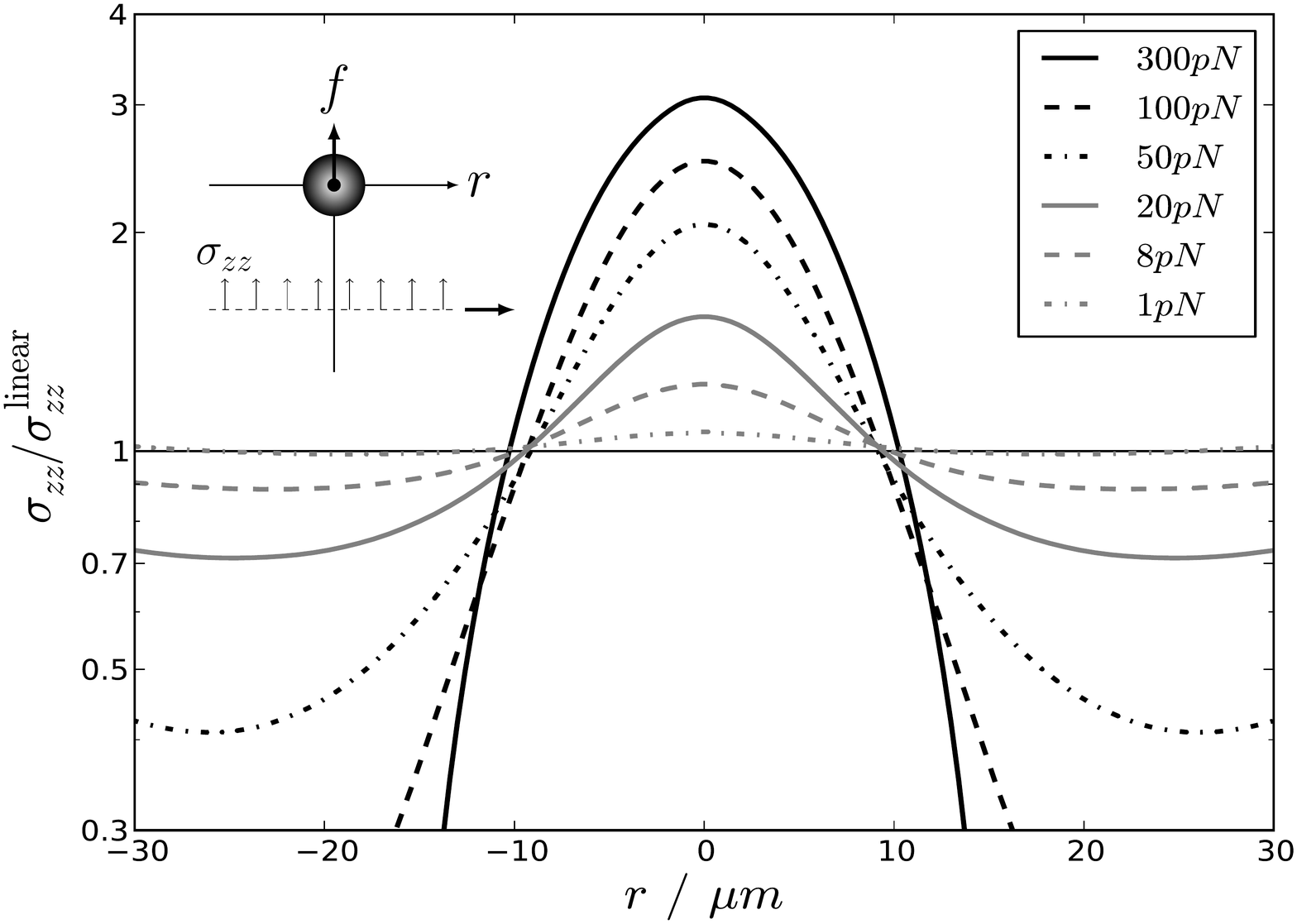}}
\caption{Stress focussing for $\ell_{\rm c}=0.55$ $\mu m$, $\ell_{\rm p}=0.5$ $\mu m$ and $\rho=16.25$ $\mu m^{-2}$. The normal component of the stress tensor parallel to the force and 10 $\mu m$ behind it, as denoted in the inset, is plotted scaled to the corresponding linear solution. The 6 applied forces are shown in the legend in the same order as the centre of the response curves from top to bottom.}
\label{f:stressLine}
\end{figure}

The non-linear force-extension curve of the wormlike chain model is responsible for the phenomenon of {\em stress focussing}, where the stresses directly behind the force are enhanced. As evident in Fig.~\ref{f:stressLine}, the normal stresses across a plane normal to the direction of the force, and directly behind it, are increased manyfold with respect to the equivalent linear solution. Conversely, stresses decrease and fall below the linear solution at greater lateral distances. Since the net force on any closed surface enclosing the monopole must balance~${\bf f}^{\rm ext}$, an increase in stress in one region is expected to be compensated by a corresponding decrease elsewhere. What is not trivial here is the location of the increase, which is a consequence of the marked strain stiffening under tension.

\begin{figure}
\centerline{\includegraphics[width=9cm]{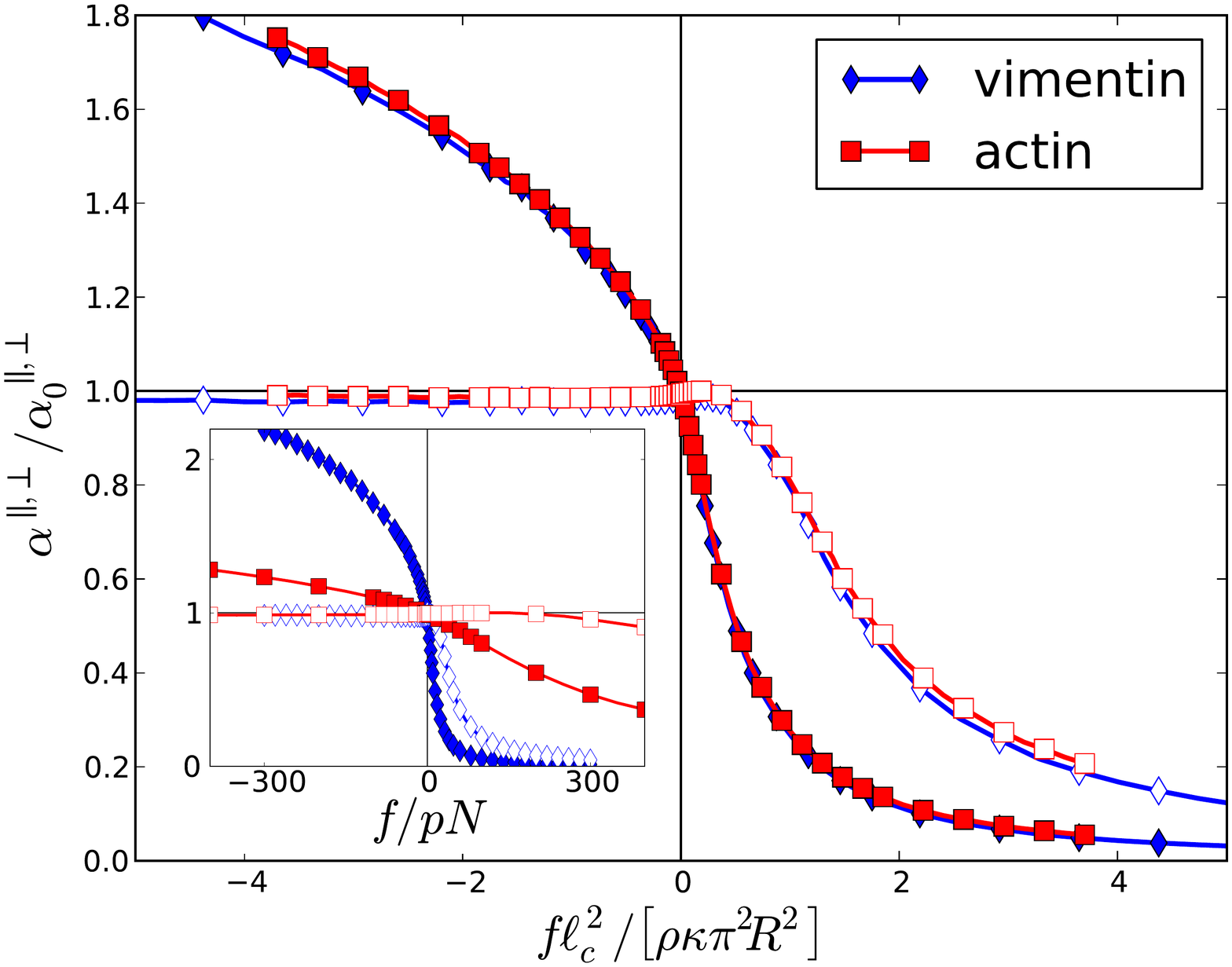}}
\caption{The local response functions at a distance 10$\mu$m behind a force monopole $f$ in parallel $\alpha^{\parallel}$ (closed symbols) and perpendicular $\alpha^{\perp}$ (open symbols) directions. Negative forces correspond to locations in front of the force. The response functions are scaled by their respective $f=0$ values, and the force is scaled by $\ell_{c}^{2}/(\rho\kappa\pi^{2}R^{2})$. The inset shows the same data plotted versus the unscaled force. Parameters were $\ell_{p}=17\mu$m, $\ell_{c}=2.2\mu$m and $\rho=39\mu{\rm m}^{-2}$ for actin~\cite{Morse1998}, and $\ell_{p}=0.5\mu$m, $\ell_{c}=0.6\mu$m and $\rho=5\mu{\rm m}^{-2}$ for vimentin~\cite{Lin2010}.}
\label{f:actin_vimentin}
\end{figure}
\begin{figure}[htbp]
\centerline{\includegraphics[width=9cm]{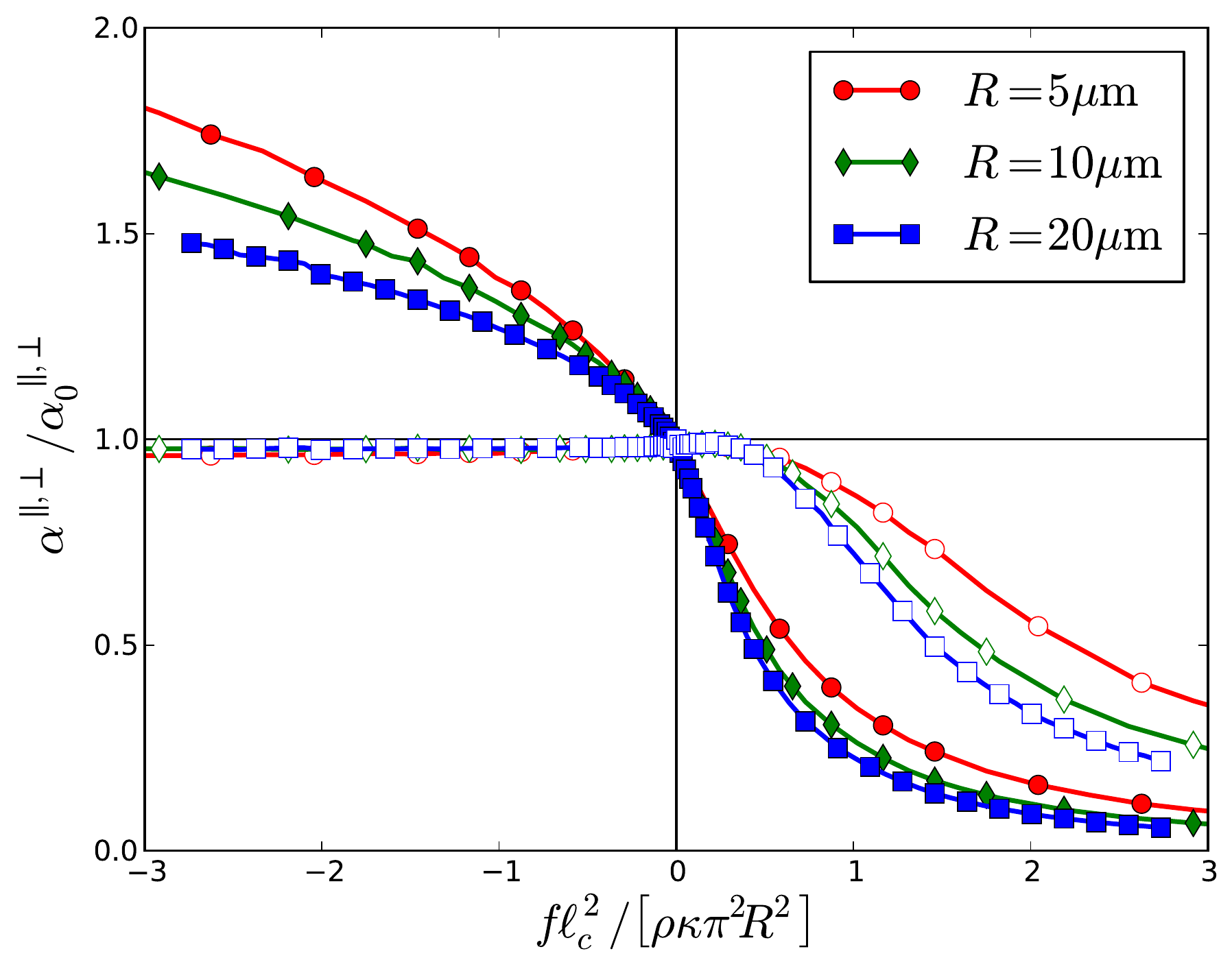}}
\caption{Normalised response curves for the vimentin parameters of Fig.\ref{f:actin_vimentin}, and $R=5\mu$m, $10\mu$m and $20\mu$m, with closed symbols for $\alpha^{\parallel}$ and open symbols for~$\alpha^{\perp}$.}
\label{f:varR}
\end{figure}

Examples of $\alpha^{\parallel,\perp}$ are given in the inset of Fig.~\ref{f:actin_vimentin}, which shows plots of the response functions $10\mu$m from the applied force for network parameters chosen to be representative of actin~\cite{Morse1998} and vimentin~\cite{Lin2010}. The softer vimentin network is clearly much more strongly affected for any given force $f$ that the stiffer actin network, as expected. It is convenient to present these results in normalised forms that can be generalised to a broader range of networks. To this end, note that, for the wormlike chain model~(\ref{e:wlc3}), the affine, hypoelastic constitutive equation (\ref{e:sigma_ij}) can be rewritten in terms of the normalised tension and extension as
\begin{equation}
\sigma_{ij}^{*}
=
\big\langle \hat{n}_{i}\hat{n}_{j}\phi(\hat{n}_{k}\hat{n}_{l}u^{*}_{kl}) \big\rangle_{\hat{\bf n}}\:,
\label{e:wlc_norm}
\end{equation}
where the normalised tension $\phi$ is the inverse of the normalised extension $u$ in~(\ref{e:wlc3}). Here, the normalised and dimensionless starred tensors are
\begin{eqnarray}
\sigma^{*}_{ij}&=&\frac{\ell_{c}^{2}}{\rho\kappa\pi^{2}}\sigma_{ij}\:,\\
u^{*}_{ij} &=& \frac{\ell_{p}}{\ell_{c}}u_{ij}\:.
\end{eqnarray}
Then the (Cartesian) force balance equations for a force monopole, $\partial_{i}\sigma_{ij}+f^{\rm ext}_{j}\delta({\bf x})=0$, become
\begin{equation}
0
=
\partial_{i}\sigma^{*}_{ij}
+
f^{\rm ext}_{j}\frac{\ell_{c}^{2}}{\rho\kappa\pi^{2}}\delta({\bf x})
\label{e:partial}
\end{equation}
which suggests a partial normalisation for the external force. To make it dimensionless, we incorporate the distance $R$ from the applied force monopole to the point at which the response functions are measured, 
\begin{equation}
f^{*}
=
\frac{f^{\rm ext}\ell_{c}^{2}}{\rho\kappa\pi^{2}R^{2}}\:.
\label{e:norm_fext}
\end{equation}
Replotting the response functions for the different network parameters (but the same $R$) against this quantity shows clear collapse as demonstrated in the main figure of Fig.~\ref{f:actin_vimentin}, confirming the validity of~(\ref{e:partial}). This means that it is only necessary to provide curves for different distances $R$ from the applied force; once the response functions are normalised to their zero prestress values~$\alpha^{\parallel,\perp}$, which are given in (\ref{e:alpha0}) (using known expressions for $G$~\cite{Lin2010}), and the force scaled as just described, the response functions are fully specified. Varying $R$ therefore generates a family of curves that can be used for fitting purposes to estimate unknown parameters. To this end, we present in Fig.~\ref{f:varR} curves for $R=5\mu$m, $10\mu$m and $20\mu$m. The shapes of these curves depend on the non-linearity and we have not been able to derive a simple scaling with $R$ that collapses them. Some empirical observations are mentioned in the discussion.

The predictions of this model have been compared to crosslinked vimentin networks driven by a colloidal particle in an optical trap~\cite{HeadSub}, in particular the response curves Fig.~\ref{f:actin_vimentin}. Qualitative agreement was found for increasing the magnitude of the driving force, but the model was found to overestimate the increase in stiffness by a factor of around~2.6, which was interpreted as due to the breakdown of the affine assumption. Interestingly, this factor was constant over the assayed range, and applying the same value consistently agreed with data for independent metrics, suggesting non-affinity may be amenable to this model with minimal modifications. Both the data and the corresponding model predictions are described in full elsewhere~\cite{HeadSub}.


\section{Discussion}
\label{s:disc}

The relevance of our findings to actual networks will depend on the likelihood that internal or external forces can be of sufficient magnitude to place the constituent fibers into their non-linear response regime, which is problem dependent. For microrheology experiments such forces can always (in principle) be reached, but they may also occur naturally. As an instructive example, taking typical actin parameters to be $\ell_{p}=17\mu$m and $\ell_{c}=2.2\mu$m~\cite{Morse1998}, the unit normalised tension $\tau_{0}\ell_{c}^{2}/\kappa\pi^{2}=1$, when non-linearities will occur, corresponds to an actual tension $\tau_{0}\approx0.14$pN. This is well within the range of forces capable of being generated in physiological conditions, such as by a single mysoin-II molecular motor~\cite{HowardBook}, confirming the relevance of non-linear fiber response to acto-myosin mixtures~\cite{Silva2011}. It is expected that similar arguments will suggest the relevance of non-linear mechanical fiber response in a broader range of networks and problems.

Future work could aim to reduce the reliance of the various assumptions that were made to close the equations, albeit at the likely expense of an increase in complexity. Relaxing the hypoelastic assumption to allow finite strains should be possible following established methods~\cite{BowerBook,OgdenBook}, and would allow direct comparison to data for collagen scaffolds used in tissue engineering~\cite{Meng2012,Tronci2013}, which exhibit a characteristic J-shaped curve for strains up to 100\%. Fitting the model to such curves will allow quantities related to the network structure to be extracted. Relaxing the affine assumption is more problematic. Given the prohibitive number of degrees of freedom required to simulate 3D networks spanning the same length scales as presented here, an alternative analytical framework is desirable, but so far none has yet been devised even for the linear response, although there has been recent progress~\cite{Zaccone2011}.

Finally, we remark on an observation that we do not yet have an explanation for. The induced response curves in Figs.~\ref{f:actin_vimentin} and~\ref{f:varR} were demonstrated to collapse after scaling the external force as per~(\ref{e:norm_fext}). However, the factor $R^{2}$ was chosen purely to make the force dimensionless. It appears that, for the range of $R$ studied, these curves can also be collapsed by scaling by $R$ as $f\ell_{c}^{2}/(R^{5/3}\rho\kappa\pi^{2})$, but {\em only for $f>0$}; the $f<0$ regime remains distinct. This piecewise collapse, which may well be approximate, presumably stems from the solution to the non-linear elasticity problem, but we have been unable to discern any simple reason for the $R$ exponent of $\approx5/3$, neither what other length scale may be included to make the final quantity dimensionless. Further work to elucidate this observation would aid in the fitting of theory to numerical data, as it would mean that only a single curve for one value of $R$ would need to be evaluated to determine the full range of axisymmetric non-linear response.

\appendix*
\section{Estimate of strain gradients due to a point force}

To estimate the strain gradient at a given distance from an applied force, we employ the known response due to a point monopole in a linear isotropic body, in the understanding this will likely be an overestimate if strain stiffening is present. The rank-3 tensor $\partial_{k}u_{ij}$ evaluated at a point ${\bf r}=r\hat{\bf r}$ relative to an external point force ${\bf f}=f\hat{\bf n}$ is found by twice differentiating the displacement field~\cite{LandauLifshitz},
\begin{widetext}
\begin{equation}
\partial_{k}u_{ij}
=
\frac{f}{16\pi\mu(1-\nu)}
\frac{1}{r^{3}}
\left\{
\begin{array}{c}
\delta_{ij}\hat{n}_{l}
\left[
\delta_{kl}-3\hat{r}_{k}\hat{r}_{l}
\right]
-3\hat{n}_{l}
\left[
\delta_{ik}\hat{r}_{j}\hat{r}_{l}
+\delta_{jk}\hat{r}_{i}\hat{r}_{l}
+\delta_{lk}\hat{r}_{i}\hat{r}_{j}
-5\hat{r}_{i}\hat{r}_{j}\hat{r}_{k}\hat{r}_{l}
\right]
\\
-(1-2\nu)
\left[
\hat{n}_{j}(\delta_{ik}-3\hat{r}_{i}\hat{r}_{k})
+\hat{n}_{i}(\delta_{jk}-3\hat{r}_{j}\hat{r}_{k})
\right]
\end{array}
\right\}
\end{equation}
\end{widetext}
where the prefactor includes the shear modulus $\mu$ and the Poisson ratio $\nu$. To determine the characteristic magnitudes, we consider all non-zero components at a point in front of the force $\hat{r}=\hat{n}$, and in the perpendicular direction with $\hat{\bf n}\cdot\hat{\bf r}=0$. Explicit evaluation reveals that, of all these, the largest in magnitude is $\partial_{\hat{r}}u_{\hat{r}\hat{r}}=f/2\pi\mu r^{3}$ evaluated in front of the force.

The assumption of affinity is expected to break down before forces $f$ for which this largest strain gradient exceeds the inverse crosslink length $\ell_{c}^{-1}$, or $f\sim \mu r^{3} \ell_{c}^{-1}$. (It may of course break down much sooner than this, including in the linear regime~\cite{Head2005}). Taking values representative of actin networks, {\em i.e.} $\ell_{p}=17\mu$m, $\ell_{c}=2.2\mu$m and $\rho=39\mu{\rm m}^{-2}$ (so $\mu\approx25$Pa)~\cite{Morse1998}, this estimate suggests a breakdown in affinity by the time $f/r^{3}\sim 10$ $pN/\mu m^{3}$, or for forces of 10pN at a range of 1$\mu$m. The same calculations for vimentin networks~\cite{Lin2010} suggests a lower force of just 1pN for the same distance.


\end{document}